\documentclass[a4paper, 12pt]{article}

\usepackage{graphicx}
\usepackage{epstopdf}
\usepackage{amsmath}
\usepackage{subcaption}
\usepackage{caption}
\usepackage[round, sort]{natbib}
\usepackage{color}
\usepackage{bm}
\usepackage{authblk}

\newcommand{\dd}{\mathrm{d}}

\let\oldequation\equation
\let\endoldequation\endequation

\renewenvironment{equation}
  {\begin{footnotesize}\oldequation}
  {\endoldequation\end{footnotesize}}

\title{Drop deformation with soluble surfactants in linear flows: role of adsorption-desorption}
\author{P. Regazzi\thanks{Email address for correspondence: paul.regazzi@univ-amu.fr}\hspace{5pt} and M. Leonetti\thanks{Email address for correspondence: marc.leonetti@univ-amu.fr}}
\affil{Aix Marseille Univ, CNRS, CINaM, Marseille, France.}

\begin{document}
\maketitle
\begin{abstract}
Drop deformation in shear flow is determined up to second order theory in $Ca$ while considering kinetic effects on surfactants distributions in steady state. Surfactants inside the drop are adsorbed faster than those on the surface  leading to an increase in total surfactants concentration on the semi-minor axis of the ellipsoidal droplet. New expressions for Taylor deformation and angle deviation are proposed involving kinetics effects. Theoretical calculations lead to a possibility of lesser deformation than the usual limit condition and a minor correction for the deviation angle.
\end{abstract}
\section{Introduction}
Rheology of fluids was first introduced by a work from \cite{taylor1932viscosity} that sought to study pure water drop deformation and breakup. It brought significant results in that field especially by quantitatively defining the difference between the main axis $L$ and the smaller one $S$ being $D_T = \frac{L-S}{L+S}$. A generalization in non stationary state was given by \cite{cox1969deformation} which takes into account mechanical stress applied on the drop surface and describes its long-term equilibrium regime in shear flow and hyperbolic flow. Deformation and breakup of drops in dilute emulsion expressions were determined by \cite{frankel1970constitutive} and \cite{barthes1973deformation}.

Later on, applications on more complex drops involving surfactants concentration evolution such as foam or red blood cells were investigated. Movements of surfactants inside and at the membrane of the drop counterbalance the surface tension, maximizing the deformation: this is the so-called Marangoni effect. It was widely explained by \cite{stone1990effects} and experimentally observed in \cite{stone1994dynamics} who considered advective-diffusive effects of surfactants in non-stationary flows and led to an increase of the Taylor deformation.

Another effect on surfactant-free droplets was first introduced by \cite{flumerfelt1980effects} and widely studied by \cite{narsimhan2019shape} involving surface viscosity effects when dealing with soluble surfactants hence increasing or decreasing Taylor deformation according to the range of values of the Boussinesq numbers $Bq_{\mu,\kappa}$.

Subsequently, rheology of fluids applied to other objects such as capsules and vesicles has proven to be worth the detour in the soft matter domain with experimental studies (tumbling and tank-treading in works from\cite{dupire2012full}, \cite{kaoui2009red}), theoretical studies given by \cite{keller1982motion}) and numerical studies by \cite{gounley2016influence}. However, the physics of surfactants' distribution and motion involved in such objects and droplets is much more complicated to observe and describe.

Surfactants are known to provide a significant contribution to the surface tension and tend to influence the ambient flow. A work from \cite{vlahovska2009small} considered the convective effect of surfactants at the surface of the drop and shew that it tends to counterbalance the viscosity in the Taylor deformation and even cancel the viscosity contribution at first order perturbation theory.

All these effects contributed to the Taylor deformation. However it seems that no matter the hypothesis made on the values for these quantities, $D_T$ is determined by the conditions at infinity of the considered flow and isn't diminished by the aforementioned effects. Questions arise such as which phenomena could cause a lower $D_T$ value  than what is the usual result.

In this paper, following the work of \cite{wailliez2024drop}, we consider kinetic effects on surfactants; the drop contains surfactants inside moving towards the interface and the ones at the surface can move towards the center. Effects of desorption and adsorption will be thus considered in order to check if it could contribute to a lower Taylor deformation in shear flow. Section \ref{sec:problem_statement}  gives a global overview of the problem using tensor calculus. Section \ref{sec:solutions} highlights expressions for Taylor deformation and angle deviation of the drop. The results are then discussed in section \ref{sec:discussion}.
\section{Problem statement}\label{sec:problem_statement}
Consider a spherical droplet of radius $R$ containing a fluid of viscosity $\eta^*$ and soluble surfactants. The drop is immersed in another fluid of viscosity $\eta$. The mechanical equilibrium of strengths applied on the drop will lead to the drop deformation. Surfactants inside the droplet are considered to be free as they can be adsorbed or desorbed from the interface. From now on, implicit summation over repeated indices is taken as convention. Let $\left( x_1, x_2, x_3\right)$ be Cartesian coordinates centered on the droplet.

\subsection{Main equations}

The radius of the droplet being at the order of the micrometer in experiments, the Reynolds number in the Navier-Stokes equation is small $Re \ll 1$. Consequently, the fluids' motion is well-described by the Stokes equation:
\begin{equation}
\begin{split}
\partial_j^2 v_i &= \frac{1}{\eta}\partial_i P,\qquad \partial_i v_i = 0\\
\partial_j^2 v_i^* &= \frac{1}{\eta^*}\partial_i P,\qquad \partial_i v_i^* = 0
\end{split}
\end{equation}
where the superscript $*$ denotes all the quantities inside the droplet,\\ $\partial_j = \partial/\partial x_j$ and $P$ is the pressure. The undisturbed velocity speed is
\begin{equation}
v_i^\infty = \dot\epsilon\left( E_{ij}+\Omega_{ij}\right) x_j\ \text{for}\ r = \sqrt{x_a x_a}\rightarrow \infty
\end{equation}
$E_{ij}$ is the symmetric rate-of-strain tensor, $\Omega_{ij}$ is the antisymmetric vorticity tensor and $\dot\epsilon$ is the shear rate. This paper will be focused on steady state solutions obtained by continuity at the interface at a distance $R$ from the center of the drop. The main equations are velocity continuity and mechanical equilibrium
\begin{equation}
v_i-v_i^* = 0\ \text{for}\ r=R\label{velocity_continuity}
\end{equation}
\begin{equation}
\left( \eta\Pi_{ij} - \eta^*\Pi_{ij}^*  \right) n_j = F_i\ \text{for}\ r=R\label{stress_continuity}
\end{equation}
where $\Pi_{ij}$ is the hydrodynamical stress tensor, $n_j$ is normal to the interface and $F_i$ are the interfacial forces given by
\begin{equation}
F_i = -\gamma P_{jk}\partial_k P_{ij} - P_{ij}\partial_j \gamma + P_{jk}\partial_k\Sigma_{ij}^*\ \text{for}\ r=R
\end{equation}
The first term on the right hand-side is the stress due to the drop's curvature and comes from a paper submitted by \cite{taylor1932viscosity}. The second term is the Marangoni stress proposed by a work from \cite{stone1990effects}. And the third term (see \cite{narsimhan2019shape}) comes from the surface viscosity $\eta_\mu$ and dilational viscosity $\eta_\kappa$ effect on total stress. Boussinesq numbers are deduced $Bq_{\kappa,\mu} = \eta_{\kappa,\mu}/\eta R$. Hence
\begin{equation}
\Sigma_{ij}^* = \left( -Bq_\kappa + Bq_\mu \right)  P_{ij}P_{lm}\partial_m v_l^* -Bq_\mu P_{ik}\left( P_{kl}\partial_l v_n^*+P_{nl}\partial_l v_k^* \right) P_{nj}
\end{equation}
\subsection{Evolution equations}

When submitted to the far-field flow, we assume that the flow lines will be deformed along the drop. As such the steady state evolution equation for the deformation is
\begin{equation}
v_i n_i = v_i^* n_i = 0\ \text{for}\ r=R
\end{equation}

For surfactants concentration inside the drop, the flux vector is
\begin{equation}
j_i = -D \partial_i C + C v_i^*,\qquad j_i n_i = -K_1 C\left( 1-\frac{\Gamma}{\Gamma_\infty} \right)+K_2 \Gamma\label{eq_flux}
\end{equation}
$D$ is the diffusion coefficient inside the drop, $K_1$ is the kinetic adsorption coefficient and $K_2$ the kinetic desorption coefficient. $\Gamma_\infty$ is the saturated surfactant concentration at the interface. Thus one also obtain the equation
\begin{equation}
v_i^* \partial_i C - D \partial_i^2 C = 0
\end{equation}
Solving such an equation implies a harmonic decomposition for $C$.
 The evolution equation becomes, much alike in \cite{stone1990effects},
\begin{equation}
P_{ij}\partial_j\left[ P_{ik} \Gamma v_k^* - \hat{D} P_{kl}\partial_l \Gamma \right] + \Gamma \left( P_{ij}\partial_j n_i\right) \left( v_k^*n_k \right) = K_1 C\left( 1-\frac{\Gamma}{\Gamma_\infty} \right) - K_2 \Gamma\label{g_evolution}
\end{equation}
with $\hat{D}$ the diffusion coefficient at the interface.
\subsection{Small-deformation analysis}
We suppose the drop shape will be slightly deformed. As the capillary number arises in such a way that $Ca = \eta \dot\epsilon R/\bar{\gamma} \ll 1$ since experimental drops have small radii, it is quite natural to consider that all the quantities involved in the equations, especially the drop deformation quantity tensor $f\left( \frac{x_1}{r},\frac{x_2}{r},\frac{x_3}{r}\right)$ can be expanded in $Ca$. For instance
\begin{equation}
R = 1 + Ca f^{(1)}+Ca^2\left[ -A f^{(1)}f^{(1)} + f^{(2)} \right]+O\left( Ca^3\right)
\end{equation}
\begin{equation}
\frac{\Gamma}{\bar{\Gamma}} =1+\Gamma'= 1 + Ca\ g^{(1)}+Ca^2\ g^{(2)}+O\left( Ca^3\right)
\end{equation}
\begin{equation}
\frac{C}{\bar{C}} = 1 + Ca\ h^{(1)}+Ca^2\ h^{(2)}+O\left( Ca^3\right)
\end{equation}
with $A$ a constant determined using the volume conservation of the drop and the bar denoting the quantity at equilibrium. The normal vector also becomes
\begin{equation}
n_i = \frac{x_i}{r}-Ca\ \partial_i f^{(1)}+o\left( Ca^2\right)
\end{equation}

Using the same method as in a work from \cite{stone1990effects} the surface tension is influenced by the following state equation
\begin{equation}
\hat{\gamma} = \gamma_0 + \hat{E}\ln\left( 1-\psi \right),\qquad \psi = \frac{\Gamma}{\Gamma_\infty}
\end{equation}
where $\gamma_0$ is the surface tension without surfactants, $\hat{\gamma}$ the surface tension at the interface and $\hat{E}$ the Gibbs elasticity. In the linear regime the equation becomes
\begin{equation}
\hat{\gamma} = \gamma_0 - \hat{E}\psi
\end{equation}
and the equilibrium state is
\begin{equation}
\bar{\gamma} = \gamma_0 - \hat{E}\bar{\psi},\qquad \bar{\psi} = \frac{\bar{\Gamma}}{\Gamma_\infty}
\end{equation}
Using such a result and multiplying equation (\ref{stress_continuity}) by $1/\eta \dot\epsilon R$ the mechanical equilibrium equation becomes
\begin{equation}
\left( \Pi_{ij} - \lambda\Pi_{ij}^*  \right) n_j = -\frac{1-\hat{E}\bar{\psi}\Gamma}{Ca\bar{\gamma}} P_{jk}\partial_k P_{ij} + \frac{\hat{E}\bar{\psi}}{Ca\bar{\gamma}}P_{ij}\partial_j \Gamma' + P_{jk}\partial_k\Sigma_{ij}^*
\end{equation}
with $\lambda = \eta^*/\eta$ the viscosity contrast.
To recover \cite{stone1990effects} notations, one has
\begin{equation}
\beta = \frac{\hat{E}}{\gamma_0}\bar{\psi}
\end{equation}
By defining the following non dimensional quantities
\begin{equation}
\zeta = \frac{\bar{\gamma} R}{\eta \hat{D}},\qquad k_1^* = K_1 \frac{R}{\hat{D}},\qquad k_2^* = K_2 \frac{R}{\hat{D}},\qquad \alpha = R \frac{\bar{C}}{\bar{\Gamma}}
\end{equation}
The flux equations (\ref{eq_flux}) and the evolution equation (\ref{g_evolution}) writes
\begin{equation}
j_i = -\frac{D}{\hat{D}}\alpha \partial_i \frac{C}{\bar{C}} + Ca \zeta \alpha \frac{C}{\bar{C}} v_i^*,\qquad j_i n_i = -k_1^* \alpha \frac{C}{\bar{C}}\left( 1-\psi \right)+k_2^* \frac{\Gamma}{\bar{\Gamma}}
\end{equation}
The evolution equation for dimensionless surfactants is
\begin{equation}
P_{ij}\partial_j\left[ Ca \zeta P_{ik} \frac{\Gamma}{\bar{\Gamma}} v_k^* -  P_{il}\partial_l \frac{\Gamma}{\bar{\Gamma}} \right] + Ca\zeta\frac{\Gamma}{\bar{\Gamma}} \left( P_{ij}\partial_j n_i\right) \left( v_k^*n_k \right) = k_1^* \frac{C}{\bar{C}}\left( 1-\psi \right) - k_2^* \frac{\Gamma}{\bar{\Gamma}}
\end{equation}

\section{Solutions}\label{sec:solutions}
Since we are dealing with the Stokes equations in a spherical drop, we introduce solid harmonics. The tensors used will require to be symmetrical and traceless in order to satisfy the Laplace equation and allow the solid harmonics decomposition
\subsection{First order}
To determine the expression of the velocities, we'll use one main result from \cite{lamb1924hydrodynamics}
\begin{equation}
v_i = \varepsilon_{ijk}\partial_j (\chi x_k)+\partial_i \phi+\frac{n+3}{2(n+1)(2n+3)}r^2\partial_i P - \frac{n}{(n+1)(2n+3)}P x_i
\end{equation}
With $n$ the solid harmonic's order. As shown by \cite{cox1969deformation}, the first harmonics that appear in the aforementioned flows are solid harmonics of degree 2. Using the same decomposition as in the work from \cite{frankel1968motion} we define
\begin{equation}
\begin{array}{lll}
P^{(1)} = T_{lm}^{(1)}\partial_l\partial_m r^{-1},& \phi^{(1)} = S_{lm}^{(1)}\partial_l\partial_m r^{-1},& \chi^{(1)} = C_l^{(1)} \partial_l r^{-1}\\\\
P^{*,(1)} = T_{lm}^{*,(1)}r^5\partial_l\partial_m r^{-1},& \phi^{*,(1)} = S_{lm}^{*,(1)}r^5\partial_l\partial_m r^{-1},& \chi^{*,(1)} = C^{*,(1)}_lr^3 \partial_l r^{-1}\\\\
f^{(1)} = F_{lm}^{(1)}r^5\partial_l\partial_m r^{-1},& g^{(1)} = G_{lm}^{(1)}r^5\partial_l\partial_m r^{-1},& h^{(1)} = H_{lm}^{(1)}r^3\partial_l\partial_m r^{-1}
\end{array}
\end{equation}
To simplify the equations, we set $r=1$.
Such decompositions give the following quantities
\begin{equation}
v_i^{(1)}= 6S_{ji}^{(1)}x_j-15 S_{lm}^{(1)}x_mx_lx_i r^{-7}+\frac{3}{2}T_{lm}^{(1)}x_lx_mx_i-\varepsilon_{ijk}C_j^{(1)} x_k+3\varepsilon_{ijk}C_l^{(1)} x_lx_kx_j
\end{equation}
\begin{equation}
v_i^{*,(1)} = 6S_{ji}^{*,(1)}x_j+\frac{5}{7}T_{ji}^{*,(1)}x_j-\frac{2}{7}T_{lm}^{*,(1)}x_lx_mx_i-\varepsilon_{ijk} C_j^{*,(1)} x_k
\end{equation}
\begin{equation}
\begin{split}
\Pi_{ij}^{(1)} =& 3 T_{lj}^{(1)}x_lx_i+3 T_{li}^{(1)}x_lx_j+12 S_{ij}^{(1)}-60S_{li}^{(1)}x_lx_j-60S_{lj}^{(1)}x_lx_i\\
&-30S_{lm}^{(1)}x_lx_m\delta_{ij}+210S_{lm}^{(1)}x_lx_mx_ix_j-15T_{lm}^{(1)}x_lx_mx_ix_j
\end{split}
\end{equation}
\begin{equation}
\begin{split}
\Pi_{ij}^{*,(1)} =& -\frac{25}{7} T_{lm}^{*,(1)}x_lx_m\delta_{ij}+12S_{ij}^{*,(1)}+\frac{10}{7}T_{ij}^{*,(1)}+\frac{6}{7}T_{li}^{*,(1)}x_lx_j+\frac{6}{7}T_{lj}^{*,(1)}x_lx_i
\end{split}
\end{equation}
\begin{equation}
\begin{split}
\Sigma_{ij}^{(1)} =& Bq_\kappa\left[ \left( 6 S_{lm}^{*,(1)}+\frac{9}{7}T^{*,(1)}\right)\delta_{ij}x_lx_m+\left( -6 S_{lm}^{*,(1)}-\frac{9}{7}T_{lm}^{*,(1)} \right) x_lx_mx_ix_j \right]\\
&+Bq_\mu\left[ -12 S_{ij}^{*,(1)}-\frac{10}{7}T_{ij}^{*,(1)}+\left( 12 S_{lj}^{*,(1)}+\frac{10}{7}T_{lj}^{*,(1)} \right) x_lx_i\right.\\
&+ \left( 12 S_{li}^{*,(1)}+\frac{10}{7}T_{li}^{*,(1)} \right) x_lx_j+\left( -6 S_{lm}^{*,(1)}-\frac{5}{7}T_{lm}^{*,(1)} \right) x_lx_m\delta_{ij}\\
&\left.+\left( -6 S_{lm}^{*,(1)}-\frac{5}{7}T_{lm}^{*,(1)} \right) x_lx_mx_ix_j \right]
\end{split}
\end{equation}
By using the continuity equations one obtains
\begin{equation}
\frac{1}{5}T_{ij}^{(1)}-2S_{ij}^{*,(1)}-\frac{1}{5}T_{ij}^{*,(1)}+\frac{E_{ij}}{3} = 0
\end{equation}
\begin{equation}
\varepsilon_{ijk}C_k^{(1)}-\varepsilon_{ijk}C_k^{*,(1)}+\Omega_{ij} = 0
\end{equation}
\begin{equation}
\begin{split}
&-\frac{3}{5}T_{ij}^{(1)}-\lambda\left( 4S_{ij}^{*,(1)}+\frac{2}{5}T_{ij}^{*,(1)} \right)+\frac{2}{3}E_{ij}+\left( Bq_\kappa - Bq_\mu \right)\left( -\frac{6}{35}T_{ij}^{*,(1)}-\frac{4}{5}S_{ij}^{*,(1)} \right)\\
&+Bq_\mu\left( -\frac{26}{35}T_{ij}^{*,(1)}-\frac{28}{5}S_{ij}^{*,(1)} \right) = \frac{8}{5}F_{ij}^{(1)}+\frac{2}{5}\frac{\hat{E}\bar{\psi}}{\bar{\gamma}}G_{ij}^{(1)}
\end{split}
\end{equation}
\begin{equation}
-9S_{ij}^{(1)}+\frac{3}{2}T_{ij}^{(1)}-6S_{ij}^{*,(1)}-\frac{3}{7}T_{ij}^{*,(1)}+E_{ij} = 0
\end{equation}
\begin{equation}
-3 \varepsilon_{ijk}C_k^{(1)} = 0
\end{equation}
\begin{equation}
\begin{split}
&72S_{ij}^{(1)}-9T_{ij}^{(1)}-\lambda\left( 12S_{ij}^{*,(1)}-\frac{3}{7}T_{ij}^{*,(1)} \right)+2e_{ij}+\left( Bq_\kappa - Bq_\mu \right)\left( \frac{18}{7}T_{ij}^{*,(1)}+12S_{ij}^{*,(1)} \right)\\
&+Bq_\mu\left( \frac{18}{7}T_{ij}^{*,(1)}+12S_{ij}^{*,(1)} \right) = 12F_{ij}^{(1)}-6\frac{\hat{E}\bar{\psi}}{\bar{\gamma}} G_{ij}^{(1)}
\end{split}
\end{equation}
where the well-know angular integral relations on the sphere
\begin{equation}
\int x_i x_j\dd\Omega = \frac{4\pi}{3}\delta_{ij},\qquad \int x_ix_jx_lx_m\dd\Omega = \frac{4\pi}{15}\left( \delta_{ij}\delta_{lm}+\delta_{il}\delta_{jm}+\delta_{im}\delta_{jl} \right)
\end{equation}
\begin{equation}
\int x_ix_jx_lx_mx_px_q\dd\Omega = \frac{4\pi}{105}\left( \delta_{ij}\delta_{lm}\delta_{pq}+14\ \text{permutations} \right)
\end{equation}
were used.

Since the Taylor deformation has already been dealt with in the literature for surfactants' concentration without kinetic effects or Boussinesq numbers, this section only develops theory including the former.
According to the expression of $h^{(1)}$, equation (\ref{eq_flux}) leads to
\begin{equation}
\left( -6\frac{D}{\hat{D}}\alpha H_{li}^{(1)}x_l+\zeta\alpha v_i^{*,(1)}\right) x_i = -3k_1^*\alpha H_{lm}^{(1)}x_lx_m+3k_2^*G_{lm}^{(1)}x_lx_m
\end{equation}
Thus
\begin{equation}
H_{li}^{(1)}x_lx_i = \frac{3k_2^* G_{lm}^{(1)}x_lx_m-\zeta\alpha v_i^{*,(1)}x_i}{\alpha\left( -6\frac{D}{\hat{D}}+3k_1^* \right)}
\end{equation}
Injecting this result in equation (\ref{g_evolution}) one finally has
\begin{equation}
\begin{split}
&P_{ij}\partial_j\left[ \zeta P_{ik} v_k^{*,(1)} -  3P_{kl}\partial_l \left( G_{pq}^{(1)}x_px_q\right) \right] + 2\left( v_k^{*,(1)}n_k^{(0)} \right)\\
& = k_1^* \frac{3k_2^* G_{lm}^{(1)}x_lx_m-\zeta\alpha v_i^{*,(1)}x_i}{-6\frac{D}{\hat{D}}+3k_1^* } - 3k_2^* G_{pq}^{(1)}x_px_q,\qquad r=R=1
\end{split}
\end{equation}
After using evolution equations,
\begin{equation}
-9 S_{ij}^{(1)} + \frac{3}{2}T_{ij}^{(1)} + E_{ij} = 0
\end{equation}
\begin{equation}
\zeta\left( -\frac{3}{7}T_{ij}^{*,(1)}-2 S_{ij}^{*,(1)} \right)+6 G_{ij}^{(1)} = k_1^*\frac{3k_2^* G_{ij}^{(1)}-\alpha\zeta\left( 6 S_{ij}^{*,(1)}+\frac{3}{7}T_{ij}^{*,(1)} \right)}{-6\frac{D}{\hat{D}}+3k_1^*} - k_2^* G_{ij}^{(1)}
\end{equation}
we have
\begin{equation}
F_{ij}^{(1)} = \frac{5}{24}\frac{\left( 19\lambda+16+8Bq_\mu+24Bq_\kappa\right)+4\zeta\frac{\hat{E}\bar{\psi}}{\bar{\gamma}}\left( 1+\Lambda \right)}{\left( 5\lambda+5+4Bq_\mu+6Bq_\kappa \right)+\zeta\frac{\hat{E}\bar{\psi}}{\bar{\gamma}}\left( 1+\Lambda \right)}E_{ij}
\end{equation}
\begin{equation}
G_{ij}^{(1)} = \frac{1}{3}\frac{5\left( 1+\Lambda\right)}{\left( 5\lambda+5+4Bq_\mu+6Bq_\kappa \right)+\zeta\frac{\hat{E}\bar{\psi}}{\bar{\gamma}}\left( 1+\Lambda \right)}E_{ij}
\end{equation}
where $\Lambda$ is the kinetic part and is written
\begin{equation}
\Lambda = \frac{1+\frac{k_1^*}{k_2^*} \bar{\psi}\alpha}{-\frac{6}{k_2^*} + 3\frac{k_1^*}{k_2^*}\frac{\hat{D}}{D}\left( 1-\bar{\psi}\right) - \left( 1+\alpha\frac{k_1^*}{k_2^*}\bar{\psi}\right)}
\end{equation}
These results can be adapted to the flow used accordingly with the rate-of-strain and vorticity tensor.
In simple shear flow one has
\begin{equation}
E_{ij} = \frac{1}{2}\begin{pmatrix}
0 & 1 & 0\\
1 & 0 & 0\\
0 & 0 & 0\\
\end{pmatrix},\qquad \Omega_{ij} = \frac{1}{2}\begin{pmatrix}
0 & 1 & 0\\
-1 & 0 & 0\\
0 & 0 & 0\\
\end{pmatrix}
\end{equation}
Thus, $f^{(1)}$ becomes
\begin{equation}
f^{(1)} = \frac{5}{8}\frac{\left( 19\lambda+16+8Bq_\mu+24Bq_\kappa\right)+4\zeta\frac{\hat{E}\bar{\psi}}{\bar{\gamma}}\left( 1+\Lambda \right)}{\left( 5\lambda+5+4Bq_\mu+6Bq_\kappa \right)+\zeta\frac{\hat{E}\bar{\psi}}{\bar{\gamma}}\left( 1+\Lambda \right)}xy
\end{equation}
In spherical coordinates, the Taylor deformation is then
\begin{equation}
D_T =  \frac{Ca}{2}\left( f^{(1)}(\phi=\frac{\pi}{4})-f^{(1)}\left(\phi=\frac{3\pi}{4}\right) \right)
\end{equation}
\begin{equation}
\frac{D_T}{Ca} = \frac{5}{16}\frac{\left( 19\lambda+16+8Bq_\mu+24Bq_\kappa\right)+4\zeta\frac{\hat{E}\bar{\psi}}{\bar{\gamma}}\left( 1+\Lambda \right)}{\left( 5\lambda+5+4Bq_\mu+6Bq_\kappa \right)+\zeta\frac{\hat{E}\bar{\psi}}{\bar{\gamma}}\left( 1+\Lambda \right)}, \qquad \phi_T = \frac{\pi}{4},\qquad \theta = \frac{\pi}{2}
\end{equation}
In the limit $\zeta\frac{\hat{E}\bar{\psi}}{\bar{\gamma}}=0,\ Bq_\mu=Bq_\kappa=0$, we recover the result in \cite{taylor1932viscosity}. In the limit $\zeta\frac{\hat{E}\bar{\psi}}{\bar{\gamma}}=0$, the deformation determined by \cite{narsimhan2019shape} is recovered.

\subsection{Second order}
According to \cite{cox1969deformation}, harmonics of order 2 and 4 are involved. We then define, still using Frankel decomposition, the quantities
\begin{equation}
\begin{array}{ll}
P^{(2)} = T_{lm}^{(2)}\frac{\partial^2 r^{-1}}{\partial x_l\partial x_m},& \phi^{(2)} = S_{lm}^{(2)}\frac{\partial^2 r^{-1}}{\partial x_l\partial x_m}\\\\
P^{(2)} = T_{lmpq}^{(2)}\frac{\partial^2 r^{-1}}{\partial x_l\partial x_m\partial x_p \partial x_q},& \phi^{(2)} = S_{lmpq}^{(2)}\frac{\partial^2 r^{-1}}{\partial x_l\partial x_m\partial x_p \partial x_q}\\\\
P^{*,(2)} = T_{lm}^{*,(2)}r^5\frac{\partial^2 r^{-1}}{\partial x_l\partial x_m},& \phi^{*,(2)} = S_{lm}^{*,(2)}r^5\frac{\partial^2 r^{-1}}{\partial x_l\partial x_m}\\\\
P^{*,(2)} = T_{lmpq}^{*,(2)}r^9\frac{\partial^2 r^{-1}}{\partial x_l\partial x_m\partial x_p\partial x_q},& \phi^{*,(2)} = S_{lmpq}^{*,(2)}r^9\frac{\partial^2 r^{-1}}{\partial x_l\partial x_m\partial x_p\partial x_q}\\\\
f^{(2)} = F_{lmpq}^{(2)}r^5\frac{\partial^2 r^{-1}}{\partial x_l\partial x_m\partial x_p\partial x_q},& g^{(2)} = G_{lmpq}^{(2)}r^5\frac{\partial^2 r^{-1}}{\partial x_l\partial x_m\partial x_p\partial x_q}
\end{array}
\end{equation}
\begin{equation*}
h^{(2)} = H_{lmpq}^{(2)}r^5\frac{\partial^2 r^{-1}}{\partial x_l\partial x_m\partial x_p\partial x_q}
\end{equation*} 
Since we consider a small perturbation of the drop shape, the quantities are Taylor expanded around $r=1+Ca f^{(1)}$. For instance the external velocity becomes
\begin{equation}
\left. v_i\right|_{r=1+Ca f^{(1)}} = \left[ Ca\ v_i^{(1)}+Ca^2\left( v_i^{(2)} + f^{(1)} x_k \partial_k v_i^{(1)} \right)+o\left( Ca^3\right)\right]_{r=1}
\end{equation}
The usual continuity equations and evolution equations are then
\begin{equation}
v_i^{(2)}-v_i^{*,(2)} + f^{(1)}x_k\partial_k\left( v_i^{(1)}-v_i^{*,(1)} \right) = 0
\end{equation}

\begin{equation}
\begin{split}
&\left( \Pi_{ij}^{(2)} - \lambda\Pi_{ij}^{*,(2)}  \right) n_j^{(0)}+f^{(1)} x_k \partial_k \left( \Pi_{ij}^{(1)} - \lambda\Pi_{ij}^{*,(1)}  \right) n_j^{(0)}+\left( \Pi_{ij}^{(1)} - \lambda\Pi_{ij}^{*,(1)}  \right) n_j^{(1)} \\
&= -\frac{1}{Ca} \left( P_{jk}^{(1)}\partial_k P_{ij}^{(1)}\right) - \frac{\hat{E}\bar{\psi}}{Ca\bar{\gamma}}\left( P_{ij}^{(0)}\partial_j \Gamma'^{,(2)}+ P_{ij}^{(1)}\partial_j \Gamma'^{,(1)}\right)\\
&+ P_{jk}^{(0)}\partial_k\Sigma_{ij}^{*,(2)}+ P_{jk}^{(0)}\partial_k\left( f^{(1)}x_m\partial_m \Sigma_{ij}^{*,(1)}\right)+ P_{jk}^{(1)}\partial_k\Sigma_{ij}^{*,(1)}
\end{split}
\end{equation}

\begin{equation}
v_i^{(2)}n_i^{(0)}+f^{(1)}x_k\partial_k v_i^{(1)}n_i^{(0)} + v_i^{(1)}n_i^{(1)} = 0
\end{equation}

\begin{equation}
\begin{split}
&\zeta P_{ij}^{(0)}\partial_j\left( P_{ik}^{(0)} v_k^{*,(2)} +P_{ik}^{(0)}f^{(1)}x_m\partial_m v_k^{*,(1)} +P_{ik}^{(0)} \Gamma'^{,(1)} v_k^{*,(1)}+P_{ik}^{(1)} v_k^{*,(1)}\right)\\
&+\zeta P_{ij}^{(1)}\partial_j\left( P_{ik}^{(0)}v_k^{*,(1)} \right) -  P_{ij}^{(0)}\partial_j\left( P_{il}^{(0)}\partial_l \Gamma'^{,(2)}+P_{il}^{(1)}\partial_l \Gamma'^{,(1)} \right) - P_{ij}^{(1)}\partial_j\left( P_{il}^{(0)}\Gamma'^{,(1)}\right)\\
& + \zeta \left( P_{ij}^{(0)}\partial_j n_i^{(0)}\right) \left( v_k^{*,(2)}n_k^{(0)}+f^{(1)}x_m\partial_m v_k^{*,(1)}n_k^{(0)}+v_k^{*,(1)}n_k^{(1)} \right)\\
&+\zeta\left( P_{ij}^{(0)}\partial_j n_i^{(1)}+P_{ij}^{(1)}\partial_j n_i^{(0)}\right)\left( v_k^{*,(1)}n_k^{(0)}\right)+\zeta \Gamma'^{,(1)}\left( P_{ij}^{(0)}\partial_j n_i^{(0)}\right)\left( v_k^{*,(1)}n_k^{(0)}\right)\\
&= k_1^* \left( \frac{C}{\bar{C}}^{(2)}-\frac{C^{(1)}\bar{C}}\psi \right) - k_2^* \Gamma'^{,(2)}
\end{split}
\end{equation}
The more detailed expressions for velocities and stress tensors are given in appendix \ref{app:A}. The angle becomes
\begin{equation}
\begin{split}
\phi = \frac{\pi}{4}&- Ca\frac{\left( 19\lambda + 16 + 24 Bq_\kappa + 8 Bq_\mu \right)\left( 2\lambda+3 \right)\left( 19\lambda+16\right)}{16\left[ 19\lambda+16+24 Bq_\kappa + 8 Bq_\mu + 4 \zeta\frac{\hat{E}\bar{\psi}}{\bar{\gamma}}\left( 1+\Lambda \right) \right]\left[ 5\lambda+5+6 Bq_\kappa + 4 Bq_\mu + \zeta\frac{\hat{E}\bar{\psi}}{\bar{\gamma}}\left( 1+\Lambda\right) \right]}\\
&- Ca\frac{4 \zeta\frac{\hat{E}\bar{\psi}}{\bar{\gamma}}\left( 1+\Lambda\right)\left[ 2\left( 2\lambda+3\right)\left( 19\lambda+16\right) + \frac{\zeta}{3}\left( \lambda+4\right)\left( 1+\Lambda\right)\right]}{16\left[ 19\lambda+16+24 Bq_\kappa + 8 Bq_\mu + 4 \zeta\frac{\hat{E}\bar{\psi}}{\bar{\gamma}}\left( 1+\Lambda \right) \right]\left[ 5\lambda+5+6 Bq_\kappa + 4 Bq_\mu + \zeta\frac{\hat{E}\bar{\psi}}{\bar{\gamma}}\left( 1+\Lambda\right) \right]}\\
&- Ca\frac{4 \zeta\frac{\hat{E}\bar{\psi}}{\bar{\gamma}}\left( 1+\Lambda\right)\left( 16 Bq_\mu^2 + \frac{\left( 240 Bq_\kappa + 8\left( 1+\Lambda\right)\zeta + 318\lambda+312\right) Bq_\mu}{3}+\frac{\left( 282\lambda+408\right) Bq_\kappa}{3}\right)}{16\left[ 19\lambda+16+24 Bq_\kappa + 8 Bq_\mu + 4 \zeta\frac{\hat{E}\bar{\psi}}{\bar{\gamma}}\left( 1+\Lambda \right) \right]\left[ 5\lambda+5+6 Bq_\kappa + 4 Bq_\mu + \zeta\frac{\hat{E}\bar{\psi}}{\bar{\gamma}}\left( 1+\Lambda\right) \right]}\\
&- Ca\frac{16 \zeta^2\frac{\hat{E}^2\bar{\psi}^2}{\bar{\gamma}^2}\left( 1+\Lambda\right)^2\left( 2\lambda+3\right)}{16\left[ 19\lambda+16+24 Bq_\kappa + 8 Bq_\mu + 4 \zeta\frac{\hat{E}\bar{\psi}}{\bar{\gamma}}\left( 1+\Lambda \right) \right]\left[ 5\lambda+5+6 Bq_\kappa + 4 Bq_\mu + \zeta\frac{\hat{E}\bar{\psi}}{\bar{\gamma}}\left( 1+\Lambda\right) \right]}
\\
&- Ca\frac{\left( 19\lambda + 16 + 24 Bq_\kappa + 8 Bq_\mu \right)\left( 46\lambda Bq_\kappa + 52 \lambda Bq_\mu + 32 Bq_\kappa Bq_\mu + 64 Bq_\kappa + 48 Bq_\mu\right)}{16\left[ 19\lambda+16+24 Bq_\kappa + 8 Bq_\mu + 4 \zeta\frac{\hat{E}\bar{\psi}}{\bar{\gamma}}\left( 1+\Lambda \right) \right]\left[ 5\lambda+5+6 Bq_\kappa + 4 Bq_\mu + \zeta\frac{\hat{E}\bar{\psi}}{\bar{\gamma}}\left( 1+\Lambda\right) \right]}
\end{split}
\end{equation}
By considering $\zeta\frac{\hat{E}\bar{\psi}}{\bar{\gamma}} = 0 = Bq_\mu = Bq_\kappa = 0$ one finds the usual result for angle deviation seen in the work from \cite{chaffey1967second}. For $\zeta\frac{\hat{E}\bar{\psi}}{\bar{\gamma}} = 0$ the result from \cite{narsimhan2019shape} is recovered.
\section{Discussion}\label{sec:discussion}

In this section we discuss drop deformation and deviation up to second order while considering kinetic effects.

\subsection{Drop deformation}

\begin{figure}
	\begin{subfigure}[t]{0.496\textwidth}
\centerline{\includegraphics[width=\textwidth]{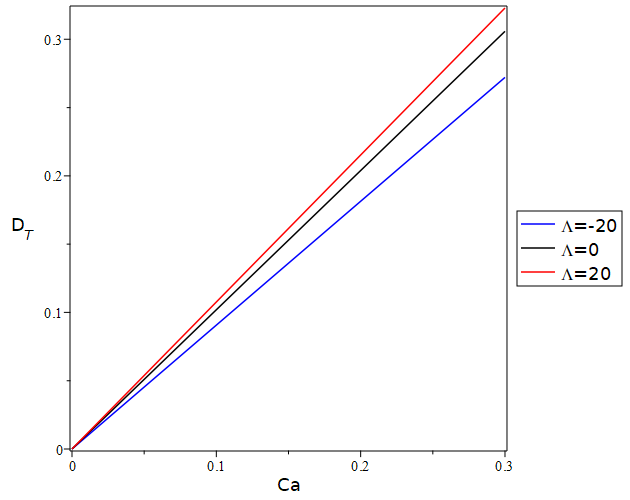}}
\end{subfigure}
	\begin{subfigure}[t]{0.496\textwidth}
\centerline{\includegraphics[width=\textwidth]{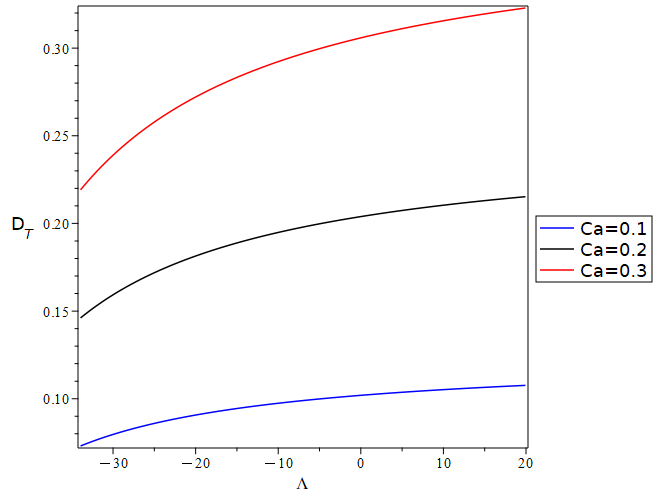}}
\end{subfigure}
\caption{Drop deformation in simple shear flow at first order theory with parameters values $\lambda=1$, $Bq_\mu = Bq_\kappa = 5$, $\zeta \hat{E}\bar{\psi}/\bar{\gamma} = 1$.}
\label{fig:dt}
\end{figure}

In shear flow, it is difficult to achieve a deformation slope $D_T/Ca$ lower than $1$ as stated by Taylor. However, the kinetic term involving $\Lambda$ has enough freedom to obtain a smaller value. Such a condition is obtained with
\begin{equation}
-\frac{\bar{\gamma}}{\zeta\hat{E}\bar{\psi}}\left( \frac{19\lambda}{4}+4+6 Bq_\kappa + 2 Bq_\mu + \zeta\frac{\hat{E}\bar{\psi}}{\bar{\gamma}}\right)\leq \Lambda<-\frac{\bar{\gamma}}{\zeta\hat{E}\bar{\psi}}\left( \frac{15\lambda}{4}+6Bq_\kappa - 6 Bq_\mu + \zeta\frac{\hat{E}\bar{\psi}}{\bar{\gamma}} \right)
\end{equation}
Characteristic values are plotted in figure \ref{fig:dt} where a slope lower than $1$ is indeed allowed for a given negative $\Lambda$ depending on parameter's values.

\begin{figure}
\centerline{\includegraphics[width=0.5\textwidth]{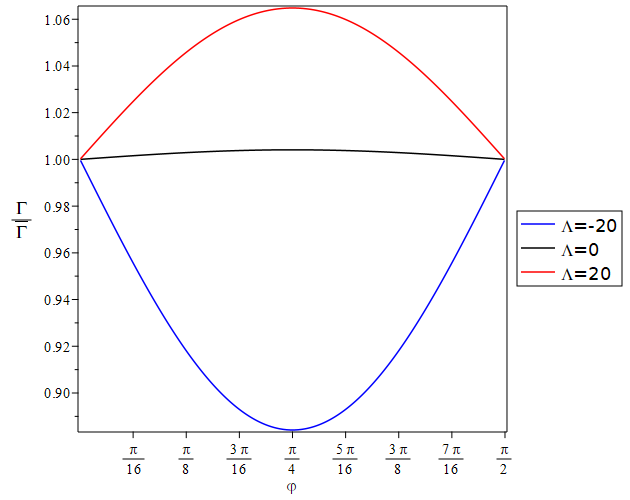}}
\caption{Dimensionless interfacial concentration of surfactants against the angle of the drop $\varphi$ with $Ca=0.1$, $\lambda=1$, $Bq_\mu = Bq_\kappa = 5$, $\zeta \hat{E}\bar{\psi}/\bar{\gamma} = 1$.}
\label{fig:gamma_phi}
\end{figure}

By looking at the expression of $\Lambda$ one finds that it may be negative if
\begin{equation}
\frac{2}{k_2^*} < \frac{D}{\hat{D}}\frac{k_1^*}{\alpha_2}
\end{equation}
which yields
\begin{equation}
2T_1 < T_D
\end{equation}
for $T_1$ the characteristic adsorption time and $T_D$ the characteristic volume diffusion time. Moreover the surface concentration is influenced by the kinetics effects leading to a decrease or increase of concentration at the extremities of the drop depending on $\Lambda$ as illustrated in figure \ref{fig:gamma_phi}.
 It implies that the surfactants move against the imposed external flow. This counter intuitive results leads to a new perspective about surfactants' distribution and should be further investigated in experiments.
\subsection{Angle deviation}
At second order, the negative values of $\Lambda$ will not play a major role in the angle deviation and will only slightly counterbalance the deviation correction. As such in figure
\begin{figure}
\begin{subfigure}[t]{0.496\textwidth}
\centerline{\includegraphics[width=\textwidth]{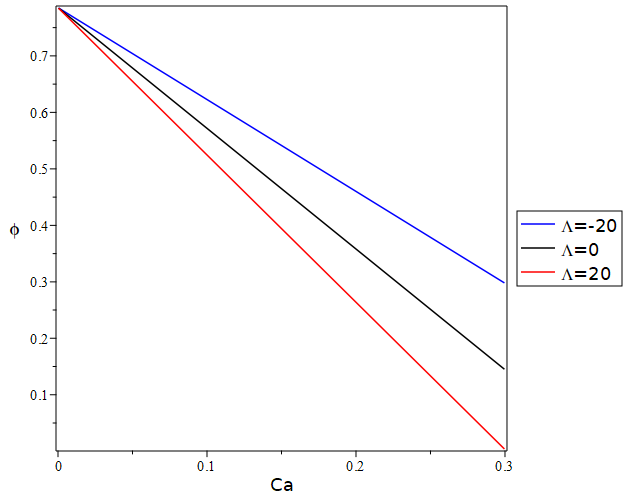}}
\end{subfigure}
\begin{subfigure}[t]{0.496\textwidth}
\centerline{\includegraphics[width=\textwidth]{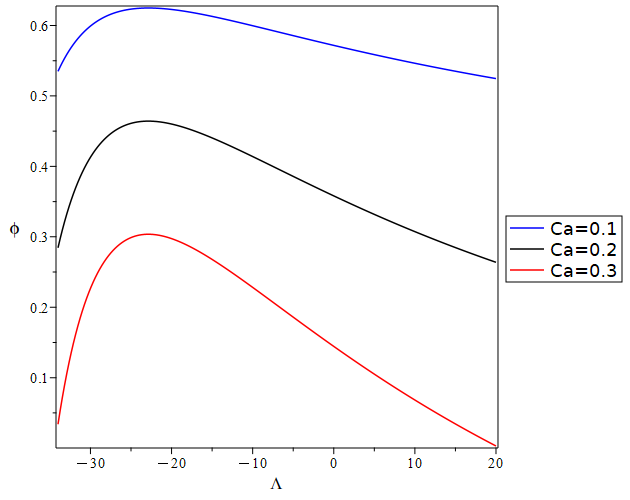}}
\end{subfigure}
\caption{Angle deviation at second order theory with $\lambda=1$, $Bq_\mu = Bq_\kappa = 5$, $\zeta \hat{E}\bar{\psi}/\bar{\gamma} = 1$.}
\label{fig:phi}
\end{figure} \ref{fig:phi} the angle would be closer to $\phi = \pi/4$. The redistribution of surfactants discussed in the previous subsection will not noticeably  influence the angle.
\section{Conclusion}
A theory for surfactants' kinetic effects was developped up to second order theory using a tensorial approach based on works from \cite{frankel1970constitutive}. The consideration of adsorption of internal surfactants and desorption of interfacial surfactants' behavior led to the definition of a quantity $\Lambda$ bringing a correction to the drop deformation and angle deviation in shear flow.

A small range of values is allowed for $\Lambda$ to bring such corrections and change the physics of surfactants' distribution depending on the viscosity ratio, Boussinesq numbers and \cite{stone1990effects} quantity. It is possible only for an adsorption characteristic time lower than the characteristic inner diffusion time.

Hence the drop's deformation allows to be smaller than the usual Taylor deformation limit condition; kinetic effects could explain some observations made in experiments \cite{} where it appears to be smaller than the common theory's prediction.

The interfacial surfactants' concentration will be higher at the semi-minor axis' location in this situation thus moving against the induced flow. An experiment checking the behavior of interfacial surfactants' concentration should be performed to further investigate this theoretical result.\newline
\newline

\qquad P. Regazzi would like to thank Y. Lebouazda (CPT, Marseille) for all his helpful discussions on the problem.
\appendix
\section{Velocities and stress tensors at second order}\label{app:A}
The velocities at second order are
\begin{equation}
\begin{split}
v_i^{(2)} =& 420 S_{lmpi}^{(2)}x_lx_mx_p - 945 S_{lmpq}^{(2)}x_lx_mx_px_qx_i-15T_{lmpi}^{(2)}x_lx_mx_p+\frac{105}{2}T_{lmpq}^{(2)}x_lx_mx_px_qx_i\\
&+6S_{ji}^{(2)}x_j-15 S_{lm}^{(2)}x_mx_lx_i +\frac{3}{2}T_{lm}^{(2)}x_lx_mx_i
\end{split}
\end{equation}
\begin{equation}
\begin{split}
v_i^{*,(2)} =& 420 S_{lmpi}^{*,(2)}x_lx_mx_p+\frac{294}{11}T_{lmpi}^{*,(2)}x_lx_mx_p-\frac{84}{11}T_{lmpq}^{*,(2)}x_lx_mx_px_qx_i\\
&+6S_{ji}^{*,(1)}x_j+\frac{5}{7}T_{ji}^{*,(2)}x_j-\frac{2}{7}T_{lm}^{*,(2)}x_lx_mx_i
\end{split}
\end{equation}
The stress tensors are
\begin{equation}
\begin{split}
\Pi_{ij}^{(2)} =& \left( 2520 S_{ijlm}^{(2)} - 90 T_{ijlm}^{(2)}\right) x_lx_m+\left( -7560 S_{jlmp}^{(2)}+315 T_{jlmp}^{(2)} \right) x_lx_mx_px_i\\
&+\left( -7560 S_{ilmp}^{(2)}+315 T_{ilmp}^{(2)} \right) x_lx_mx_px_j-1890 S_{lmpq}^{(2)} x_lx_mx_px_q\delta_{ij}\\
&+\left( 20790 S_{lmpq}^{*,(2)}-945 T_{lmpq}^{*,(2)} \right) x_lx_mx_px_qx_ix_j+3 T_{lj}^{(2)}x_lx_i+3 T_{li}^{(2)}x_lx_j\\
&+12 S_{ij}^{(2)}-60S_{li}^{(2)}x_lx_j-60S_{lj}^{(2)}x_lx_i-30S_{lm}^{(2)}x_lx_m\delta_{ij}+210S_{lm}^{(2)}x_lx_mx_ix_j\\
&-15T_{lm}^{(2)}x_lx_mx_ix_j
\end{split}
\end{equation}
\begin{equation}
\begin{split}
\Pi_{ij}^* =&\left( 2520 S_{lmij}^{*,(2)}+\frac{1764}{11}T_{lmij}^{*,(2)} \right) x_lx_m -\frac{1323}{11}+T_{lmpq}^{*,(2)}x_lx_mx_px_q\delta_{ij}+\frac{252}{11}T_{jlmp}^{*,(2)}x_lx_mx_px_i\\
&+\frac{252}{11}T_{ilmp}^{*,(2)}x_lx_mx_px_j
-\frac{25}{7} T_{lm}^{*,(2)}x_lx_m\delta_{ij}+12S_{ij}^{*,(2)}+\frac{10}{7}T_{ij}^{*,(2)}+\frac{6}{7}T_{li}^{*,(2)}x_lx_j\\
&+\frac{6}{7}T_{lj}^{*,(2)}x_lx_i
\end{split}
\end{equation}
For the \cite{narsimhan2019shape} tensor we have
\begin{equation}
\begin{split}
\Sigma_{ij} = &Bq_\kappa\left[  \left( 1260 S_{lmpq}^{*,(2)}+\frac{1050}{11}T_{lmpq}^{*,(2)} \right) x_lx_mx_px_q\delta_{ij}-\left( 1260 S_{lmpq}^{*,(2)}+\frac{1050}{11}T_{lmpq}^{*,(2)} \right) x_lx_mx_px_qx_ix_j\right.\\
&\left. +\left( 6 S_{lm}^{*,(2)}+\frac{9}{7}T^{*,(1)}\right)\delta_{ij}x_lx_m+\left( -6 S_{lm}^{*,(2)}-\frac{9}{7}T_{lm}^{*,(2)} \right) x_lx_mx_ix_j\right] \\
&+ Bq_\mu\left[ -\left( 2520 S_{lmij}^{*,(2)}+\frac{1764}{11}^{*,(2)} \right) x_lx_m+\left( 2520 S_{lmpi}^{*,(2)}+\frac{1764}{11}T_{lmpi}^{*,(2)} \right) x_lx_mx_px_j\right.\\
& +\left( 2520 S_{lmpj}^{*,(2)}+\frac{1764}{11}T_{lmpj}^{*,(2)} \right) x_lx_mx_px_i -\left( 1260 S_{lmpq}^{*,(2)}+\frac{882}{11}T_{lmpq}^{*,(2)} \right) x_lx_mx_px_q\delta_{ij}\\
&-\left( 1260 S_{lmpq}^{*,(2)}+\frac{882}{11}T_{lmpq}^{*,(2)} \right) x_lx_mx_px_qx_ix_j+-12 S_{ij}^{*,(2)}-\frac{10}{7}T_{ij}^{*,(2)}+\left( 12 S_{lj}^{*,(2)}+\frac{10}{7}T_{lj}^{*,(2)} \right) x_lx_i\\
&\left. + \left( 12 S_{li}^{*,(2)}+\frac{10}{7}T_{li}^{*,(2)} \right) x_lx_j
+\left( -6 S_{lm}^{*,(2)}-\frac{5}{7}T_{lm}^{*,(2)} \right) x_lx_m\delta_{ij}+\left( -6 S_{lm}^{*,(2)}-\frac{5}{7}T_{lm}^{*,(2)} \right) x_lx_mx_ix_j\right]
\end{split}
\end{equation}
\bibliographystyle{apalike}
\bibliography{biblio_these}
\end{document}